%% file: main.tex
\documentclass{sig-alternate}
\usepackage{mathptmx} 

\usepackage{dblfloatfix}
\usepackage{xcolor}
\usepackage{fancyhdr}
\usepackage[normalem]{ulem}
\usepackage[hyphens]{url}
\usepackage[sort,nocompress]{cite}
\usepackage[final]{microtype}
\usepackage[keeplastbox]{flushend}
\usepackage[bookmarks=true,breaklinks=true,letterpaper=true,colorlinks,linkcolor=black,citecolor=blue,urlcolor=black]{hyperref}

\fancypagestyle{firstpage}{
  \fancyhf{}
  
  \fancyfoot[C]{\thepage}
}

\usepackage{comment}
\usepackage{multirow}
\usepackage{cleveref}
\usepackage{amsmath,amssymb,amsfonts,mathtools}
\usepackage[ruled,vlined,linesnumbered]{algorithm2e}

\DontPrintSemicolon
\SetKwComment{tcp}{$\triangleright$ }{}
\SetVlineSkip{0cm}
\SetKwInOut{Param}{Param}
\SetKwInOut{Input}{Input}

\newcommand{\ourtool}{{Marvel}}

\newcommand{\TODO}[1]{\textcolor{red}{TODO: #1}}

\newcommand{\HK}[1]{\textcolor{green}{HK: #1}}

\newcommand{\betterparagraph}[1]{\noindent\textbf{#1. }}
\DeclarePairedDelimiter{\ceil}{\lceil}{\rceil}

\title{Marvel: A Data-centric Compiler for DNN Operators on Spatial Accelerators} 

\begin{document}

\numberofauthors{9} 
%
\author{
%
%
\alignauthor
Prasanth Chatarasi\\
       \affaddr{Georgia Tech}\\
       \email{cprasanth@gatech.edu}
\alignauthor
Hyoukjun Kwon\\
       \affaddr{Georgia Tech}\\
       \email{hyoukjun@gatech.com}
\alignauthor Saurabh Malik\\
       \affaddr{Georgia Tech}\\
       \email{smalik48@gatech.edu}
\and  
\alignauthor Natesh Raina\\
       \affaddr{Georgia Tech}\\
       \email{nateshraina@gatech.edu}
\alignauthor Vaisakh Haridas\\
       \affaddr{Georgia Tech}\\
       \email{vharidas3@gatech.edu}
\alignauthor Angshuman Parashar\\
       \affaddr{NVIDIA}\\
       \email{aparashar@nvidia.com}
\and
\alignauthor
Michael Pellauer\\
       \affaddr{NVIDIA}
       \email{mpellauer@nvidia.com}
\alignauthor
Tushar Krishna\\
       \affaddr{Georgia Tech}
       \email{tushar@ece.gatech.edu}
\alignauthor 
Vivek Sarkar\\
       \affaddr{Georgia Tech}
       \email{vsarkar@gatech.edu}
}

\maketitle
\thispagestyle{firstpage}
\pagestyle{plain}


\input{tex/00-Abstract}

\input{tex/01-Introduction}
\input{tex/02-Background}
\input{tex/03-Constraints}

\input{tex/04-tranformation}
\input{tex/03-Approach}
\input{tex/04-Evaluation}

\input{tex/05-Relatedwork}
\input{tex/06-Conclusion}


\bibliographystyle{IEEEtranS}
\bibliography{main}

\end{document}

%% file: tex/00-Abstract.tex
\begin{abstract}
The efficiency of a spatial DNN accelerator depends heavily on the compiler and its cost model ability to generate optimized mappings for various operators of DNN models on to the accelerator's compute and memory resources.
But, existing cost models lack a formal boundary over the operators for precise and tractable analysis, which poses adaptability challenges for new DNN operators.
To address this challenge, we leverage the recently introduced Maestro Data-Centric (MDC) notation.
We develop a formal understanding of DNN operators whose mappings can be described in the MDC notation, because any mapping adhering to the notation is always analyzable by the MDC's cost model. 
Furthermore, we introduce a transformation for translating mappings into the MDC notation for exploring the mapping space.

Searching for the optimal mappings is challenging because of the large space of mappings, and this challenge gets exacerbated with new operators and diverse accelerator configurations.
To address this challenge, we propose a decoupled off-chip/on-chip approach that decomposes the mapping space into off-chip and on-chip subspaces, and first optimizes the off-chip subspace followed by the on-chip subspace.  
The motivation for this decomposition is to reduce the size of the search space dramatically and also to prioritize the optimization of off-chip data movement, which is 2-3 orders of magnitude more compared to the on-chip data movement.
We implemented our approach in a tool called {\em Marvel}, and another major benefit of our approach is that it is applicable to any DNN operator conformable with the MDC notation.

Overall, our approach reduced the mapping space by an $O(10^{10})$ factor for the four major CNN models (AlexNet, VGG16, ResNet50, MobileNetV2), while generating mappings that demonstrate a geometric mean performance improvement of 10.25$\times$ higher throughput and 2.01$\times$ lower energy consumption compared with three state-of-the-art mapping styles from past work.
We also evaluated our approach over the GEMM, LSTM, and MLP workloads and also compared with the optimizers from past work.

\end{abstract}

%% file: tex/01-Introduction.tex
\section{Introduction}
\label{sec:introduction}

Deep learning (DL) is a fundamental technology for many emerging applications such as autonomous driving~\cite{bojarski2016end}, translation~\cite{wu2016google}, and image classification~\cite{russakovsky2015imagenet}, with accuracy close to, and even surpassing, that of humans~\cite{karpathy2015deep,toshev2014deeppose,farabet2013learning}.
%
%
Achieving low latency and energy goals with stringent computation and memory constraints of deep neural network models (DNNs) for mobile~\cite{apple_neural_core} and edge~\cite{edge_tpu} devices has emerged as an important challenge.
To cope with this challenge, specialized hardware accelerators for DNN inference are being
developed and deployed~\cite{chen2019eyeriss, DBLP:journals/micro/ChungFOPCMLLAHA18, nvdla, edge_tpu, xDNN-web}.
Most of these accelerators are ``spatial", i.e., they are built by interconnecting hundreds to thousands of processing elements (PEs). 
They achieve high throughput by exploiting parallelism over the PEs and energy efficiency by maximizing data reuse within the PE array via direct data forwarding between PEs and the use of scratchpad memories~\cite{eyeriss_isca,chen2014diannao,nvdla,parashar2017scnn,sharma2016high,jouppi2017datacenter,ma2017optimizing,zhang2015optimizing}.

\begin{figure}[!ht]
    \centering
    \includegraphics[width=\linewidth]{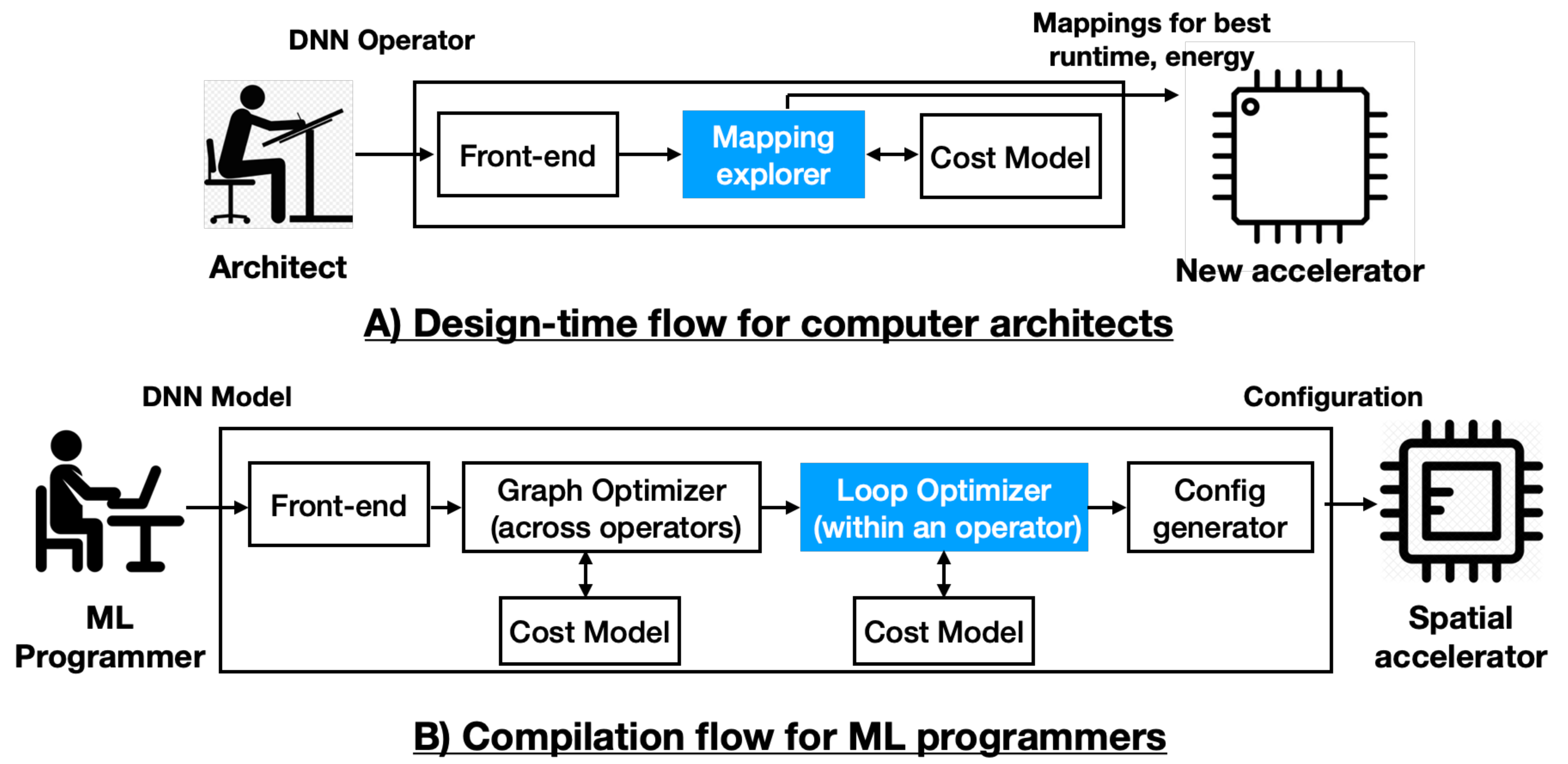}
    \caption{Overview of the design-time flow for computer architects developing new accelerators, and the compilation flow for ML programmers leveraging the accelerators. Scope of this work is the mapping explorer and the loop optimizer in the above diagram.}
    \label{fig:intro-overview}
\end{figure}

The efficiency of accelerators depends heavily on the compiler's ability to generate optimized \textit{mappings} for various operators of DNN models on to the accelerator's compute and memory resources.
%
A mapping involves parallelization, tiling, and scheduling  strategies~\cite{angshu2019timeloop,kwon2018maestro}. 
Optimized compilers (or mappers) optimizing various DNN operators are necessary during compile-time for ML programmers, and design-time for computer architects to understand reuse and data movement behaviors to design a new accelerator, as shown in ~\autoref{fig:intro-overview}. 
Thus, expressing DNN mappings and determining optimal ones is a crucial component of DNN deployment on accelerators.


Mappings are often expressed as {\em loop nests}, a syntax that resembles a simple imperative programming language with explicit parallelism.
Many cost models such as TimeLoop~\cite{angshu2019timeloop}, DMazeRunner~\cite{DMazeRunner}, Interstellar~\cite{interstellar} are developed over the loop nest description of mappings.
The loop nests syntax is very generic and can help architects/compilers in expressing a wide range of operator mappings, but the underlying cost models may not analyze all possible mappings expressible in loop nests.
Furthermore, these cost models do not have a formal boundary over DNN operators for precise and tractable analysis.
Having such no formal boundaries can bring adaptability challenges to these cost models in the compiler infrastructures and also to computer architects for design-time exploration of new DNN operators onto accelerators.

In this paper, we address the above challenge.
We leverage the recently introduced  ``Maestro Data-Centric'' (MDC) notation~\cite{kwon2018maestro} for expressing mappings.
MDC is promising because any mapping adhering to the notation can be analyzable using the MDC's cost model.
Moreover, the notation explicitly defines data mapping and organization, instead of inferring it from loop nests. 
The overall focus of this work is on (1) developing a formal understanding of DNN operators whose mappings can be described in the MDC notation, (2) introducing a transformation for translating mappings into the MDC notation for exploring the mapping space, and finally (3) proposing an efficient exploration strategy to quickly navigate the large mapping space of DNN operators. 
The key contributions are briefly described below.



{\bf 1) Conformable DNN operators.} 
The promising aspect of the MDC notation, i.e., analyzability, comes at the cost of its expressiveness. 
In this work, we introduce a formal set of rules (Section~\ref{sec:mdc-constraints}) in identifying DNN operators whose mappings can be described in the MDC notation.
We call an operator satisfying the formal rules as the {\em conformable} operator, and Table~\ref{tab:dnn-operators} lists the conformability of the popular operators with the MDC notation.



{\bf 2) Transformation.} The MDC notation is powerful in expressing and reasoning complex mappings of DNN operators onto the diverse spatial accelerators, but explicitly writing and exploring such mappings can be error-prone and tedious.
Computer architects~\cite{angshu2019timeloop} and DNN compiler frameworks~\cite{Chen:2018:TAE:3291168.3291211} view the operators and their mappings majorly in the loop nest form.
Hence, we introduce a {\em transformation} (Section~\ref{sec:compiler-transformation}) that translates a mapping specified in the loop nest form to the MDC notation and can help both the architects and compilers for mapping space exploration.

{\bf 3) Mapping space exploration.} 
The efficiency of any mapping is tightly cross-coupled with both the algorithmic aspects of DNN operators and the microarchitectural aspects of accelerators. 
Searching for the optimal mapping is challenging because of a massive space of possible loop transformations on the operators. For example, there are over 10$^{19}$ valid mappings for the CONV2D on average for mapping ResNet50~\cite{Resnet} and MobileNetV2~\cite{sandler2018mobilenetv2} on a representative DNN edge accelerator.
This challenge gets exacerbated with new operators (e.g., depth-wise) and diverse hardware accelerator configurations.
Much of the prior work~\cite{zhang2015optimizing, ma2017optimizing, zhao2019mRNA, DBLP:journals/corr/YangPRBRKRPH16} targeted hardware with limited capabilities or fixed certain aspects of the mapping space such as choice of parallel loops and loop orders~\cite{interstellar,DMazeRunner,zhang2015optimizing,ma2017optimizing,motamedi2016design}.
Approaches supporting broader classes of architectures and mappings suffer from a combinatoric explosion in the size of mapping space.

Our approach for the mapping problem is motivated by the observation that the off-chip data movement between DRAM and accelerator is 2-3 orders of magnitude more compared to the on-chip data movement involving the PE array and the local scratchpad buffers~\cite{eyeriss_isca,DBLP:journals/corr/SzeCYE17}.
Hence, we propose an approach (Section~\ref{sec:approach}) referred as ``decoupled off-chip/on-chip" that decomposes the mapping space into two subspaces, i.e., off-chip and on-chip subspaces, and first optimizes the off-chip subspace followed by exploring the on-chip mapping subspace constructed with the optimal mappings from the off-chip subspace.
In contrast to prior work~\cite{angshu2019timeloop,interstellar,DMazeRunner}, we use different approaches  and cost models for these subspaces, i.e., a classical distinct-block (DB) locality cost model~\cite{ferrante1991estimating,Sarkar:1997:ASH:271819.271828} to explore the off-chip subspace, and the MDC's cost model~\cite{kwon2018maestro} for the on-chip subspace.

We implemented the above approach in a tool called ``Marvel'', and our approach is applicable to any operator conformable with the MDC notation.
Given a conformable DNN operator, workload sizes, and a target accelerator configuration, \ourtool{}{} explores the mapping space of the operator using the decoupled approach and then outputs the mappings optimized for runtime and energy. 
Overall, our approach reduced the mapping space by an $O(10^{10})$ factor for the four major CNN models (AlexNet, VGG16, ResNet50, MobileNetV2), while generating mappings that demonstrate a geometric mean performance improvement of 10.25$\times$ higher throughput and 2.01$\times$ lower energy consumption compared with three state-of-the-art mapping styles from past work.
We also evaluated our approach over the GEMM, LSTM, and MLP workloads and also compared \ourtool{}{} generated mappings with the optimizers from past work.

%% file: tex/02-Background.tex
\section{Background}
\label{sec:background}

In this section, we provide a brief overview of the spatial DNN accelerators and also the MDC notation to describe mappings of a DNN operator onto the accelerators.

\subsection{Spatial DNN Accelerators}
\label{subsec:spatial_accelerators}

\begin{figure}[!ht]
    \centering
    \includegraphics[scale=0.2]{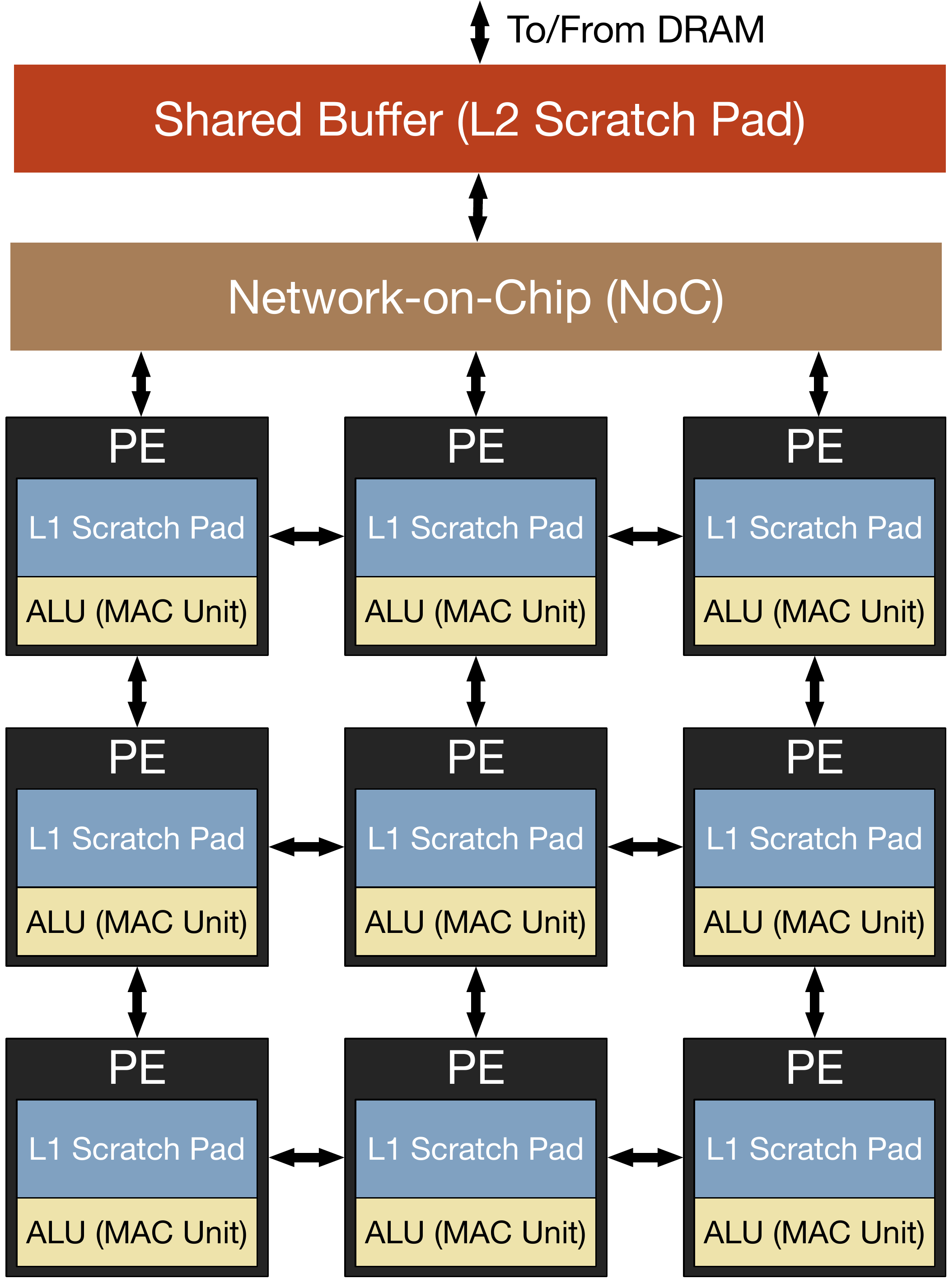}
    \caption{Abstract spatial accelerator model which is pervasive in many state-of-the-art accelerators~\cite{eyeriss_isca, nvdla, kwon2018maeri, jouppi2017datacenter}.
    }
    \label{fig:abstract_hardware_model}
\end{figure}

Spatial DNN accelerators based on ASICs and FPGAs have emerged to address extreme demands on performance and energy-efficiency of CNN layers ~\cite{eyeriss_isca,chen2014diannao, nvdla,parashar2017scnn,sharma2016high,jouppi2017datacenter}.
Such accelerators are built using an array of processing elements (PEs) to provide high parallelism and use direct communication instead of via shared memory for energy-efficiency.
An abstract model of spatial accelerators is shown in~\cref{fig:abstract_hardware_model}, where each PE of an accelerator consists of a single/multiple ALU(s) dedicated for multiply-accumulate operations (MACs) and a local scratchpad (L1 buffer).
Also, accelerators employ various network-on-chips (NoCs) for direct communication among PEs and between PE array and L2 scratchpad buffer. 
The interconnection network often supports multi-casting data to multiple PEs, which can reduce the total number of data reads from L2 buffer to PEs. 
Unlike GPU cores, PEs can communicate with adjacent PEs (data forwarding) using a NoC, which can significantly reduce the energy consumption for expensive L2 buffer accesses.
%
%
Accelerators also typically employ a large shared L2 scratchpad buffer to stage data from DRAM and also partial accumulations from PE arrays.
Both L1 and L2 scratchpad buffers are software-controlled memories, i.e., programmer/compiler directly controls contents of the buffer, unlike cache memories, which implicitly manages them, and this is because the memory traffic in accelerators is known in advance.
%
%
Many spatial accelerators can be further interconnected together to create a scale-out system
~\cite{DBLP:journals/micro/ChungFOPCMLLAHA18}.

\subsection{MDC Notation}
\label{subsec:mdc-notation}

\begin{figure}[!ht]
    \centering
    \includegraphics[width=\linewidth]{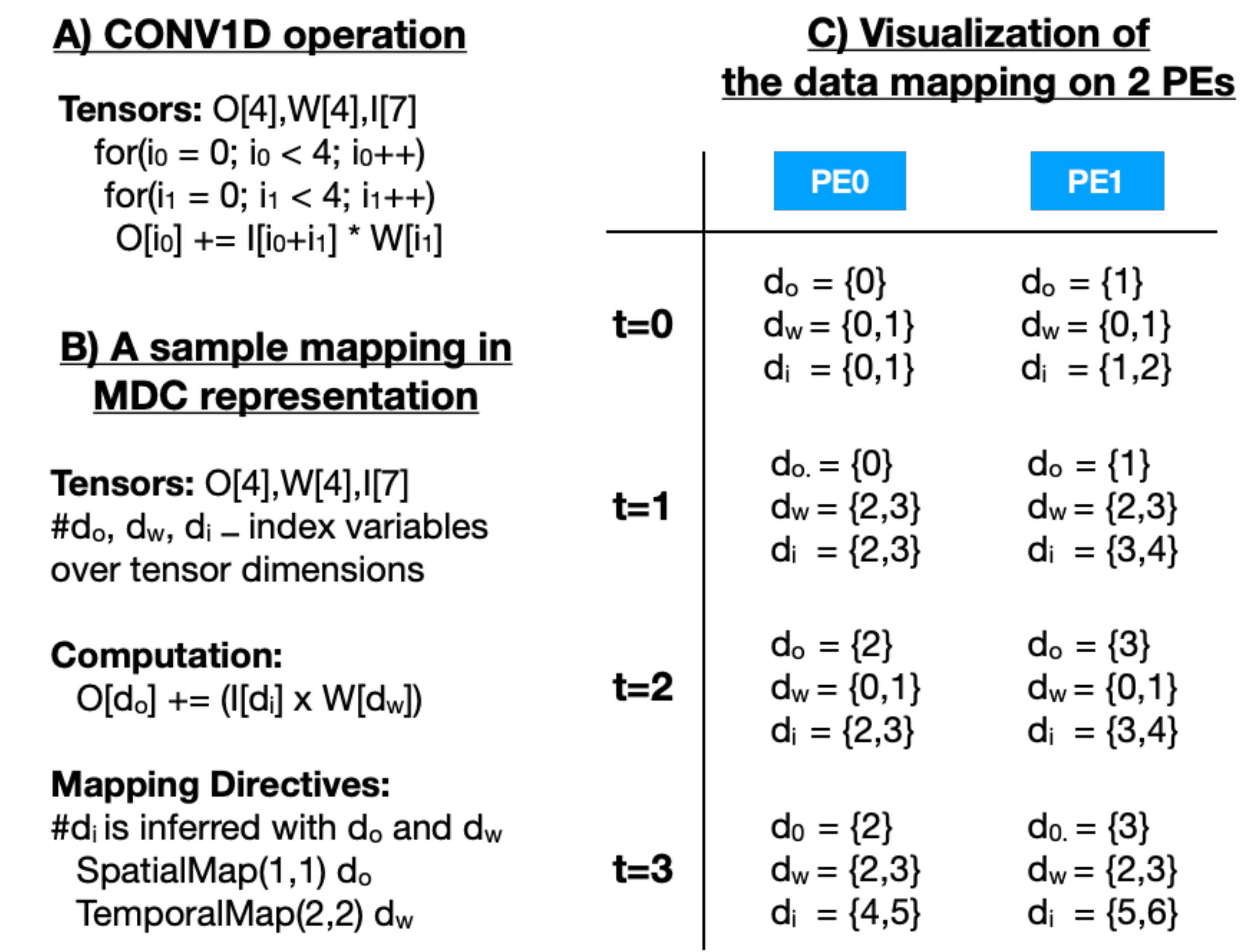}
    \caption{ A mapping of the CONV1D in the MDC notation along with the visualization of its data mappings.}
    \label{fig:mdcrepresentation}
\end{figure}

The Maestro Data-Centric (MDC) notation for a DNN operator mapping onto a spatial accelerator consists of two aspects, i.e., 1) Computation and tensor sizes, and 2) Data mapping directives over tensor dimensions.
A sample mapping of the CONV1D operator in the MDC notation is shown in~\cref{fig:mdcrepresentation}(B).
A major novelty of the MDC notation is that the data mappings of tensors across space (PEs) and time are explicitly specified using a set of data mapping directives, which makes the MDC's cost-model to estimate data movement and reuse behaviors of a mapping precisely and quickly.
We briefly describe the data mapping directives of the MDC notation with the mapping in~\cref{fig:mdcrepresentation}(B) as the example.

{\bf 1) TemporalMap (size, offset) $d$} specifies a distribution of the dimension $d$ of a tensor across time steps in a PE, and the mapped set of dimension indices is same across PEs in a given time step. 
The {\tt size} parameter refers to the number of contiguous indices mapped in the dimension $d$ to each PE, and the {\tt offset} parameter describes the shift in the starting indices of $d$ across consecutive time steps in a PE.
For instance, the directive {\tt TemporalMap(2,2) $d_{w}$} in the running example represents the distribution of first dimension ($d_{w}$) of the weight tensor with two indices mapped in each time step (i.e., $d_{w}$=\{0,1\} in PE0 and PE1 at t = 0).
Also, the offset of two denotes the increment in $d_{w}$ index after each time step  (i.e., $d_{w}$=\{2,3\} in PE0 and PE1 at t = 1) till the extent of $d_{w}$ dimension is explored.

{\bf 2) SpatialMap (size, offset) $d$} specifies a distribution of the dimension $d$ of a tensor across PEs.
The {\tt size} parameter refers to the number of contiguous indices mapped in the dimension $d$  to each PE, and the {\tt offset} describes the shift in the starting indices of $d$  across consecutive PEs.
For instance, the directive {\tt SpatialMap(1,1) $d_{O}$} in the running example represents the distribution of first dimension ($d_{O}$) of the output tensor with one index mapped to each PE (i.e., $d_{O}$=\{0\} in PE0 and $d_{O}$=\{1\} in PE1 at t = 0).
If the number of PEs is not sufficient to cover all indices of the dimension mapped, then the mapping is folded over time across the same set of PEs.

{\bf 3) Directive order.} The sequence of spatial and temporal map directives in a mapping dictates the change of data mappings to PEs across time.
Similar to a loop order, all the dimension indices corresponding to a mapping directive are explored before its outer mapping directive in the sequence begins exploring its next set of indices.
For instance, the sequence of directives in the running example, i.e., spatial map over $d_{O}$ followed by temporal map over $d_{W}$ dictates that all the dimension indices of the weight tensors need to be explored before exploring the next set of $d_{O}$ indices. 
This order results in accumulating the partial results of an output before computing another output, popularly referred to as ``output stationary'' mapping~\cite{du2015shidiannao}.
However, the sequence notation has a limitation that it cannot capture scenarios where more than one dimension index is simultaneously changing over time (except at the dimension boundaries). 
 
{\bf 4) Clusters (size)} logically groups multiple PEs or nested sub-clusters with the group size as the {\tt size} parameter.
For example, {\tt Cluster (2)} directive on an accelerator with ten PEs arranges the PEs into five clusters with the cluster size as two.
All the mapping directives above a cluster directive operate over the introduced logical clusters, while those below the cluster directive operate within a logical cluster. 
The cluster directive is extremely useful in exploiting spatial distribution of more than one tensor dimensions (e.g., row-stationary mapping~\cite{eyeriss_isca}).
Also, the directive helps in constructing hierarchical accelerators by recursive grouping.

The above aspects of the MDC notation can help in precisely specifying a wide range of mappings, including popular and sophisticated mapping styles such as row-stationary in Eyeriss~\cite{eyeriss_isca}, weight-stationary in NVDLA in~\cite{nvdla}, output-stationary in ShiDianNao~\cite{du2015shidiannao} accelerators.
However, its not clearer if all mapping behaviors of an operator can be represented in the MDC notation.


%% file: tex/03-Constraints.tex
\section{Conformable DNN Operators} \label{sec:mdc-constraints}

\begin{figure}[!t]
    \centering
    \includegraphics[width=\linewidth]{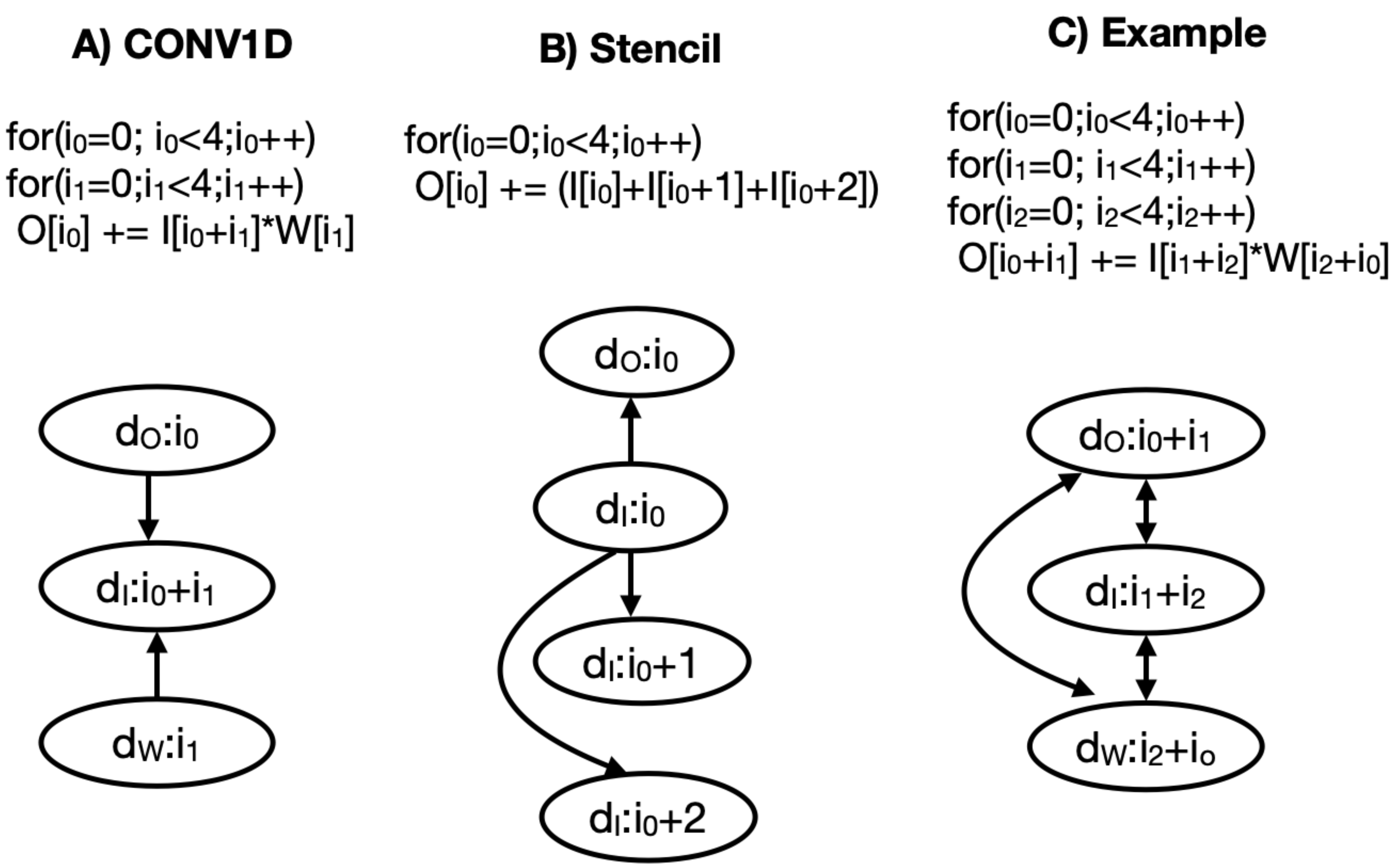}
    \caption{The dimension dependence graph (DDG) of simple operators such as CONV1D and stencil satisfying the rule R3, and an example violating the rule R3.
    d$_{O}$/d$_{I}$/d$_{W}$: tensor dimension variables corresponding to the output, input, and weight tensors.}
    \label{fig:sdg-graphs}
\end{figure}

In this section, we introduce formal rules in identifying conformable DNN operators whose mappings (reuse, parallelization and tiling strategies) can be described using the MDC notation.
We discuss rules over the abstract loop nest notation of DNN operators without any transformations for reuse and parallelization (e.g., CONV1D in~\cref{fig:sdg-graphs}).

{\bf R1: A conformable DNN operator in the abstract loop nest form must be a perfectly nested loop without any conditional statements.}

\noindent
The MDC notation restricts its computation to be uniform across all PEs at all time-steps.
This restrict is satisfied if the computation is enclosed in a perfectly nested loop without any conditional statements.
Most of the DNN operators such as CONV2D, GEMM, MLP (more in~\cref{tab:dnn-operators}) can be expressed in the form of perfectly nested loops without any conditionals. 
But, there can be implementation of certain operators such as fused convolutions, where each PE requires executing the non-uniform computation.
Hence, such operators are discarded and are non-conformable to the MDC notation.

{\bf R2: The perfectly nested loop must not have any dependences (flow, anti, output) except reduction dependences, and thus the loops can be freely reordered.}

\noindent
The MDC notation restricts the input and output tensors of an operator to be different, and this results in not having any flow and anti dependences between the tensors. 
However, the notation can support reduction operations (e.g., add, max, min), and this leads to supporting reduction dependences, i.e., flow, anti, output dependences only on the output tensor. 
Similar to the rule R1, most of the DNN operators mostly have only reduction dependences, except few operators such as parametric multi-step LSTMs which have flow dependences.

{\bf R3: The dimension dependence graph (DDG) of the perfectly nested loop must have a topological ordering, and the subscripts of dependent dimension variables of the DDG graph must be in the form of linear combination of its loop iterators.
}

\noindent
The directive order (sequence of mapping directives) of the MDC notation dictates the change of the data mappings to PEs across time.
As described in the~\cref{subsec:mdc-notation}, the directive order has limitations in capturing more than one tensor dimension variable changing simultaneously over the time (except at boundaries).
We introduce a directed graph called {\it Dimension Dependence Graph} (DDG) to find the possibility of such data movement behaviors in a DNN operator.


Each node of a DDG graph denotes a tensor dimension variable along with the array subscript referenced in that dimension. 
For instance, the node ({\tt d$_{I}$:i$_{0}$+i$_{1}$}) in~\cref{fig:sdg-graphs}(a) represents the tensor subscript {\tt i$_{0}$+i$_{1}$} used in the input tensor dimension with name {\tt d$_{I}$}.
The edges of the DDG are constructed as follows: 1) An edge is added from a node having a SIV/MIV subscript\footnote{Single Index Variable (SIV) subscript involves one loop iterator, whereas Multiple Index Variable (MIV) subscript involves more than one loop iterator~\cite{DBLP:books/mk/AllenK2001}.} to another node having a MIV subscript if there is a common loop iterator in their subscripts. 
For e.g., there is a directed edge from the node ({\tt d$_{O}$:i$_{0}$}) to ({\tt d$_{I}$:i$_{0}$+i$_{1}$}) in~\cref{fig:sdg-graphs}(a) since they have a loop iterator {\tt i$_{0}$} in common.
2) All the SIV subscripts are grouped based on their loop iterators, and then edges are added from the SIV subscript of a group having the lowest constant value (randomly choose if there exists multiple) to other SIV subscripts in the same group.
For e.g., there is a directed edge from the node ({\tt d$_{I}$:i$_{0}$}) to all the nodes ({\tt d$_{I}$:i$_{0}+1$}), ({\tt d$_{I}$:i$_{0}+2$}), and ({\tt d$_{O}$:i$_{0}$}) in~\cref{fig:sdg-graphs}(b).
3) If there is a loop iterator (say {\tt i}) dependent on other loop iterators (say {\tt j}) in its loop bounds, then construct an edge from a node with subscript having the loop iterator {\tt i} to other nodes having the loop iterator {\tt j} in their subscripts.

Now, the possibility of having multiple dimension variables changing simultaneously is reduced to the problem of finding a topological ordering in the DDG graph. In essence, the absence of a topological ordering indicates the presence of mutually dependent dimension variables (e.g., example in~\cref{fig:sdg-graphs}(c)).
In the presence of a topological ordering, the MDC notation requires the data mappings of independent dimension variables to be specified, and these variables are identified from the nodes of the DDG graph having zero in-degree.
For e.g., in the case of CONV1D in~\cref{fig:sdg-graphs}(a), only the data mappings of dimension variables related to output and weight tensors must be specified, and the dimension variable related to the input tensor is inferred by the underlying MDC's cost model.
Hence, the subscripts of dependent dimension variables need to be linear expressions of loop iterators so as to be analyzable by the MDC's cost model.
In addition, the MDC notation expects to have only one data mapping over an independent dimension variable.
If there exists more than one node with zero in-degree in the DDG graph associated with the same dimension variable, then we consider that DNN operator to be non-conformable.




{\bf R4: The subscripts associated with the independent dimension variables of the DDG graph must be in the form of linear combinations of its loop iterators with the positive unit coefficients and no constants.}

A mapping directive (either spatial or temporal) over a dimension variable restricts the variable to start from zero and increase with unit stride.
These restrictions don't allow the dimension variable to have strided increments or negative strides. 
To characterize the implication of above restrictions, we assume the abstract loop nest form of the DNN operator to be normalized, i.e., its loop iterators start from zero and have unit strides. 
To support the restricts imposed the mapping directives, each subscript (in the normalized form) associated with an independent dimension variable must be in the form of a linear combination of the subscript's loop iterators with the positive unit coefficients and no constants.
For e.g., the subscript {\tt i$_{0}$} associated with the dimension variable {\tt d$_{O}$} in~\cref{fig:sdg-graphs}(a) is in the linear form of its iterators ({\tt i$_{0}$}) with coefficient as one and no constant.

With positive unit coefficients and no constants, the SIV subscript associated with an independent dimension variable is simply an unique loop iterator (e.g., {\tt i$_{0}$} for {\tt d$_{O}$}, {\tt i$_{1}$} for {\tt d$_{W}$} in~\cref{fig:sdg-graphs}(a)).
Furthermore, the MIV subscript associated with an independent dimension variable is also in the form of adding the subscript's loop iterators. 
These loop iterators cannot be part of any subscripts associated with other dimension variables; otherwise, their in-degree wouldn't have been zero.
Hence, the loop iterators corresponding to such MIV subscript can be merged into a single loop.
{\it Overall, the subscripts associated with each of the independent dimension variables are simply unique loop iterators} (e.g., {\tt i$_{0}$} for {\tt d$_{O}$}, {\tt i$_{1}$} for {\tt d$_{W}$} in~\cref{fig:sdg-graphs}(a)).



Finally, an operator is said to MDC conformable if it satisfies all the four rules described above.
Table~\ref{tab:dnn-operators} lists the set of popular DNN operators and the conformability of these operators with the MDC notation.
As can be seen, the MDC notation can capture most of the DNN operators except parametric LSTM's, and the mappings of these operators can be analyzable by the MDC's cost model.

\begin{table}[!ht]
\centering
\scalebox{0.85}{
\begin{tabular}{|c|c|c|c|c|c|c|}
\hline
\textbf{\begin{tabular}[c]{@{}c@{}}DNN \\ Operator\end{tabular}}                 & \textbf{Types}                                                    & \textbf{R1} & \textbf{R2} & \textbf{R3} & \textbf{R4} & \textbf{\begin{tabular}[c]{@{}c@{}}Conformable \\ to MDC\end{tabular}} \\ \hline
\textbf{CONV1D}                                                                  & Regular                                                           & Y           & Y           & Y           & Y           & Y                                                                      \\ \hline
\multirow{4}{*}{\textbf{CONV2D}}                                                 & Regular                                                           & Y           & Y           & Y           & Y           & Y                                                                      \\ \cline{2-7} 
                                                                                 & \begin{tabular}[c]{@{}c@{}}Point-wise, \\ Depth-wise\end{tabular} & Y           & Y           & Y           & Y           & Y                                                                      \\ \cline{2-7} 
                                                                                 & \begin{tabular}[c]{@{}c@{}}Strided, \\ Dilated\end{tabular}       & Y           & Y           & Y           & Y           & Y                                                                      \\ \cline{2-7} 
    \hline
\textbf{MLP}                                                                     & \begin{tabular}[c]{@{}c@{}}Fully\\  connected\end{tabular}        & Y           & Y           & Y           & Y           & Y                                                                      \\ \hline
\textbf{Pooling}                                                                 & Max, Avg                                                          & Y           & Y           & Y           & Y           & Y                                                                      \\ \hline
\multirow{2}{*}{\textbf{GEMM}}                                                   & Regular                                                           & Y           & Y           & Y           & Y           & Y                                                                      \\ \cline{2-7} 
                                                                                 & Triangular                                                        & Y           & Y           & Y           & Y           & Y                                                                      \\ \hline
\multirow{2}{*}{\textbf{LSTM}}                                                   & Single cell                                                       & Y           & Y           & Y           & Y           & Y                                                                      \\ \cline{2-7} 
                                                                                 & \begin{tabular}[c]{@{}c@{}}Parametric\\ multi-cell\end{tabular}   & Y           & N           & Y           & Y           & N                                                                      \\ \hline
\multirow{2}{*}{\textbf{\begin{tabular}[c]{@{}c@{}}Element\\ wise\end{tabular}}} & Residual                                                          & Y           & Y           & Y           & Y           & Y                                                                      \\ \cline{2-7} 
                                                                                 & ReLU                                                              & Y           & Y           & Y           & Y           & Y                                                                      \\ \hline
\textbf{Stencils}                                                                & Regular                                                           & Y           & Y           & Y           & Y           & Y                                                                      \\ \hline
\end{tabular}
}
\caption{Conformability of the popular DNN operators onto the MDC notation (Y/N refers to YES/NO).}
\label{tab:dnn-operators}
\end{table}


%% file: tex/04-tranformation.tex
\section{Transformation}
\label{sec:compiler-transformation}
The MDC notation is powerful in expressing and reasoning complex mappings of DNN operators onto the diverse spatial accelerators, but explicitly writing and exploring such mappings can be error-prone and tedious.
Computer architects~\cite{angshu2019timeloop} and DNN compiler frameworks~\cite{Chen:2018:TAE:3291168.3291211} view the operators and their mappings majorly in the loop nest form~\cite{angshu2019timeloop,zhang2015optimizing,ma2017optimizing}.
This section introduces a transformation to translate a mapping of the conformable DNN operator in the loop-nest form into the MDC notation. 
In this work, we assume the target spatial accelerators having three levels of the memory hierarchy (private L1 buffer, shared L2 buffer, and DRAM). However, our transformation can be easily extendable to more levels of hierarchy.

As described in the~\cref{subsec:mdc-notation}, the MDC notation consists of two aspects, i.e., 1) Computation and tensor sizes, and 2) Data mapping directives over independent tensor dimensions. 
The statements enclosed in the perfectly nested loop form of the conformable DNN operator are used as the computation, and the tensor sizes are extracted from the workload configuration.  
The computation and tensor sizes of the MDC notation remains the same for each mapping of the operator.
Then, the dimension dependence graph of the operator is constructed to identify the set of independent tensor dimension variables (having zero in-degree).
If there are no such independent dimension variables, then the operator is discarded as non-conformable.
The rest of the section focuses on generating data mapping directives for each mapping.

\subsection{Data Mapping directives}
\begin{figure}[!t]
    \centering
    \includegraphics[width=\linewidth]{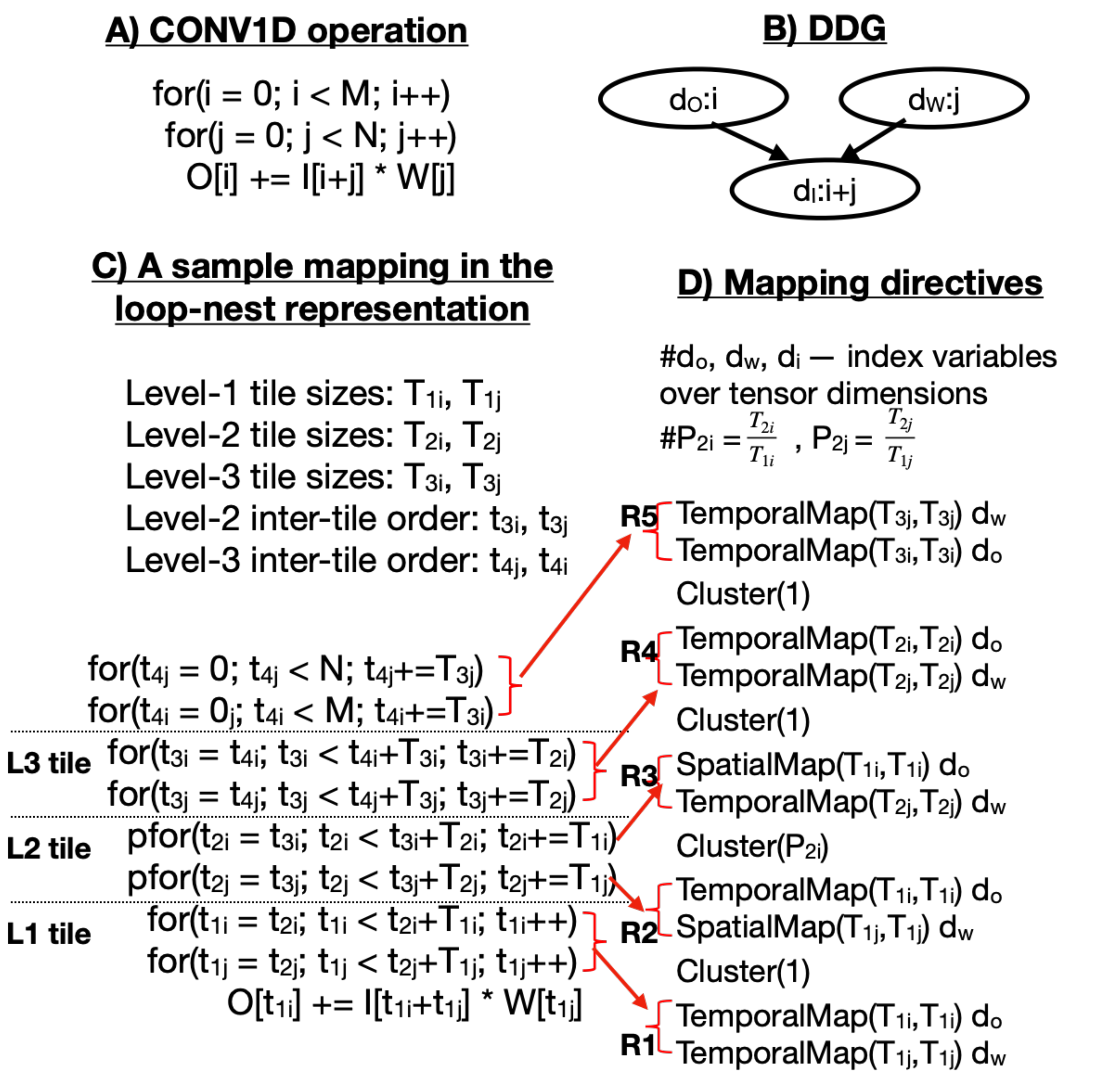}
    \caption{A brief overview of the mapping expressed in the loop-nest form of CONV1D, and its translation into the MDC notation with data mapping directives.}
    \label{fig:mapping}
\end{figure}

According to the rule R2, the loops of a conformable DNN operator can be freely reordered, so it is safe to perform multi-level tiling to exploit temporal reuse across each level of the memory hierarchy and also to exploit parallelism of the accelerator.
Each tiling, reuse and parallelization behavior of an  operator onto a spatial accelerator is referred to as a ``mapping''.
An example of the mapping of a CONV1D operation over a 3-level accelerator is shown in~\cref{fig:mapping} (C), and the different aspects of the mapping are described below.

\betterparagraph{1) Multi-level tiling tile sizes} A mapping includes tile sizes of all loop iterators for each level of tiling, i.e., 1) Level-1 tiling for the private L1 buffer, 2) Level-2 tiling for the parallelism, and 3) Level-3 tiling for the shared L2 buffer.

\betterparagraph{2) Inter-tile loop orders} A mapping also includes inter-tile loop orders\footnote{An n-dimensional loop nest after one level of tiling will have 2n loops. The outer n-loops are referred to as inter-tile loops and the later n-loops as intra-tile loops. The innermost n-loops after multi-level tiling are called as point-loops.} to describe the execution order of tiles reflecting various reuse opportunities.
E.g., the level-2 inter-tile loop order reflects spatio-temporal reuse over the PE array, and the level-3 inter-tile loop order reflects temporal reuse over the on-chip L2 buffer. 
But, the level-1 inter-tile loop order doesn't reflect any reuse, because these loops are annotated with parallelism.
Also, the loop order among point-loops doesn't provide any reuse opportunities because there is no more intermediate staging between the PE and its L1 buffer.

An n-level tiling will have {\tt n} set of tile-loops (including parallel loops) and a set of point-loops.
Each set of loops can have a different data movement (reuse) behavior based on its sizes and loop order.
We introduce a term called ``region'' to denote a sequence of data mapping directives over independent tensor dimension variables (e.g., Region R1 in~\cref{fig:mapping}(d)) without any cluster directives, and each region captures the data movement behavior present in each set of loops.
Given a mapping of the operator in the form of multi-level tile sizes and inter-tile loop orders, our approach transforms the mapping into the MDC notation as per the following steps.

\betterparagraph{1) Point-loops} As described in Rule 4, each subscript associated with an independent dimension variable is simply an unique loop iterator.
Our approach translates each loop of point-loops into a temporal map directive over the corresponding independent dimension variable with {\tt size} and {\tt offset} parameters of the directive being the point-loop size.
For, e.g., the point loop {\tt t$_{1i}$} with tile size as T$_{1i}$ in~\cref{fig:mapping}(c) is directly translated into {\tt TemporalMap(T$_{1i}$,T$_{1i}$) d$_{O}$} in the region R1 shown in~\cref{fig:mapping}(d).
Since the loop order among the point-loops doesn't provide any reuse benefits, the directive order in the region R1 doesn't matter.

\betterparagraph{2) Parallel-loops} Since each independent dimension variable is uniquely associated with a loop iterator, parallel execution of each loop iterator introduces a different data movement behavior. 
Hence, for each parallel loop, we introduce a region with a spatial map over the dimension variable associated with the parallel loop, and the temporal maps for the rest of the dimension variables in the region.
For, e.g., there are two regions with name R2 and R3 for the parallel loops corresponding to t$_{2j}$ and t$_{2i}$, respectively.
Also, the dimension d$_{W}$ associated with the iterator t$_{2j}$ and the dimension d$_{O}$ associated with the iterator t$_{2i}$ are translated into spatial maps in R2 and R3 regions respectively.
The {\tt size} and {\tt offsets} of each spatial map over a dimension variable is derived from the strides of the parallel loop iterators corresponding to the dimension variable.
The order of directives in each region corresponding to a parallel loop doesn't matter because the number of iterations arising from the rest of the temporal maps is one.
Each region corresponding to a parallel loop (except the innermost) is ended with a cluster directive with {\tt size} as the number of iterations in the parallel loop.
For, e.g., the region R3 is ended with a cluster directive with {\tt size} as the number of iterations of the loop {\tt t$_{2i}$}.

\betterparagraph{3) Inter-tile loops}
For each set of tile-loops excluding parallel loops, our transformation generates a region by creating a temporal map directive for each loop of the set with the {\tt size} and {\tt offset} of the directive as the loop stride.
For, e.g., the inter-tile loop {\tt t$_{3j}$} with stride as T$_{2j}$ in~\cref{fig:mapping}(c) is directly translated into {\tt TemporalMap(T$_{2j}$,T$_{2j}$)d$_{I}$} in the region R4 shown in~\cref{fig:mapping}(d).
The order of directives in a region is governed by the loop order among the corresponding tile-loops.
For, e.g., the level-3 inter-tile loop order ({\tt t$_{3j}$,t$_{3i}$}) dictates the temporal map over d$_{W}$ outer compared to temporal map over d$_{O}$ in region R5.
Furthermore, each region is separated by cluster directive with {\tt size} one to support different data movement behaviors across each set of tile-loops.

%% file: tex/03-Approach.tex
\section{Mapping Space Exploration}
\label{sec:approach}

The mapping space of a conformable DNN operator onto an accelerator having three levels of memory hierarchy is a cross product of valid level-1 tile sizes, level-2 tile sizes (parallelism), level-2 inter-tile loop orders, level-3 tile sizes, and level-3 inter-tile loop orders.
For example, there are over 10$^{19}$ valid mappings for a single CONV2D operator on average for mapping ResNet50 and MobileNetV2 on a representative DNN edge accelerator.
Because of this massive space of mappings, searching for efficient mappings is really challenging.
This challenge gets exacerbated with new operators (e.g., depth-wise) and diverse hardware accelerator configurations (e.g., tree-based interconnect~\cite{kwon2018maeri}).

We consider (optional) a limited form of data-layouts, i.e., innermost dimension reordering~\cite{li2016optimizing} for the tensors of operators on the DRAM.
Overall, the mapping space of an operator is a Cartesian product of six dimensions which represent different aspects of a mapping, i.e., 1) level-1 tile sizes, 2) level-2 tile sizes (parallelism), 3) level-2 inter-tile loop orders, 4) level-3 tile sizes, 5) level-3 inter-tile loop orders, and 6) data-layout of tensors.
The first three dimensions are grouped under ``{\it on-chip mapping subspace}" since they influence parallelization and on-chip data movement, and the remaining three dimensions are grouped under ``{\it off-chip mapping subspace}" since they influence the off-chip data movement.

\begin{figure}[!ht]
    \centering
    \includegraphics[width=\linewidth]{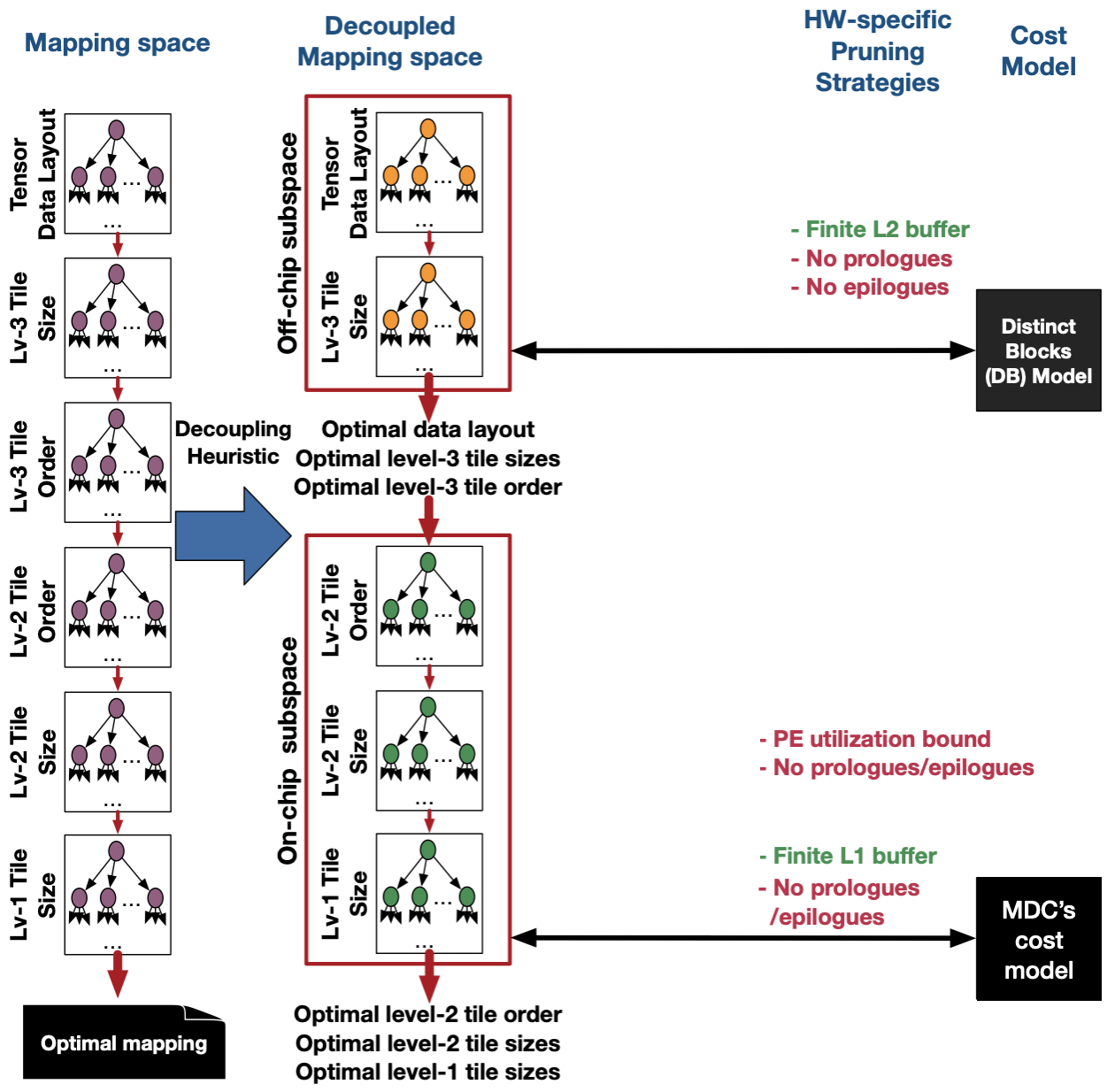}
    \caption{An overview of our approach along with pruning strategies for searching mapping space of convolutions. The pruning strategies in green color preserve optimal mappings, whereas the strategies in red color may prune optimal.}
    \label{fig:overview}
\end{figure}

Our approach towards the mapping space exploration is motivated by the observation that the off-chip data movement between DRAM and accelerator is 2-3 orders of magnitude more  compared to the on-chip data movement.
Hence, we propose an approach referred as ``decoupled off-chip/on-chip" that decomposes the mapping space into two subspaces, i.e., off-chip and on-chip subspaces, and first optimizes the off-chip subspace followed by the on-chip subspace which is constructed with the optimal mappings from the off-chip subspace.
In contrast to prior work~\cite{angshu2019timeloop,interstellar,DMazeRunner}, we use different approaches  and cost models for these subspaces, i.e., a classical distinct-block (DB) locality cost model~\cite{ferrante1991estimating,Sarkar:1997:ASH:271819.271828} to explore the off-chip subspace, and the MDC's cost model~\cite{kwon2018maestro} for the on-chip subspace.
The overall approach is implemented as a standalone tool (shown in~\cref{fig:overview}) that takes a conformable DNN operator, workload sizes, and a target accelerator configuration, then explores the mapping space of the operator using the decoupled approach, and finally outputs the mappings optimized for runtime and energy. 

\subsection{Solving off-chip mapping subspace}
\label{subsec:off-chip-dataflow-search-space}

%
The goal of finding an optimal mapping in the off-chip mapping subspace is to minimize off-chip data movement between DRAM and the L2 buffer of an accelerator. 
In our work, we assume the L2 buffer to be a software-managed scratchpad buffer, and {\it reducing the off-chip data movement\footnote{In case of non-software-managed scratchpad buffers, reducing data movement between DRAM and L2 buffer is equivalent to finding a level-3 tile whose memory footprint can fit into the L2 buffer and is maximum.} is equivalent to finding a level-3 tile that has highest arithmetic intensity}, this is because the highest arithmetic intensity results in higher reuse and less data transfer.

In our approach, we consider the classical distinct-block (DB) locality cost model~\cite{ferrante1991estimating} to measure the off-chip data movement cost, which was developed as part of the memory cost analysis to guide automatic selection of loop transformations and also optimal tile size selections~\cite{Sarkar:1997:ASH:271819.271828,Sarkar:2000:AML:1153923.1154542,Shirako:2012:ABO:2259230.2259238} in IBM XL compilers. 
The DB model is a good choice for our approach, since the model only focuses on optimizing for off-chip data movement. Moreover, it focuses only on perfectly nested loop, and conformable DNN operators are perfectly nested loops as per the rule R1 in~\cref{sec:mdc-constraints}.

The distinct blocks (DB) model starts with data-layouts of multi-dimensional arrays and also the parametric tiled version of a perfectly nested loop. Then, the model symbolically estimates the off-chip data movement cost involved in a tile of computation by measuring the number of the distinct number of DRAM blocks required for all the references in the tile of computation. 
Assuming the array {\tt I} is laid out in the row-major order, the distinct number of DRAM blocks (with block size as {\tt B} and tile sizes T$_{X}$, T$_{Y}$) required for an example array reference {\tt I[x+y][y]} enclosed in a triply nested loop with iterators {\tt x, y, z} is computed as follows:
$$DB_{I}(T_{X}, T_{Y}) \approx \Bigg( \ceil[\bigg]{\frac{T_{X} + T_{Y}}{b}} \Bigg) \times (T_{Y}) \times T_{Z}$$

%
In the above formulation, the innermost access of the reference is divided by the block size\footnote{Setting block size to one ignores the impact of data-layouts that we consider in our approach (innermost dimension reordering~\cite{li2016optimizing}).}, because the data movement with DRAM happens in multiples of block sizes. 
%
%
Now, the total data movement cost (DMC), a.k.a. memory cost per iteration, involved in a tile is computed as the number of distinct DRAM blocks required for all references in the tile by the total number of iterations in the tile. 
%
%
The optimal level-3 tile sizes and data-layouts are computed by minimizing the data movement cost function for every layout and tile sizes in the off-chip mapping subspace with the two constraints, i.e., 1) the tile size of a loop should be greater than 0 and should not exceed its corresponding loop bound, and 2) the total data required (including double buffering) for a level-3 computation tile should fit into the on-chip L2 buffer.

After computing the optimal level-3 tile sizes and data-layouts of tensors, our approach computes the partial derivatives (slopes) of the data movement cost function (based on the optimal data-layout) with respect to parametric level-3 tile sizes (similar to~\cite{Sarkar:1997:ASH:271819.271828}), and evaluate the partial derivatives by substituting optimal level-3 tile sizes.
The key insight is that having a higher negative value of a partial derivative along a loop indicates the lesser distinct number of elements referenced along the loop, i.e., highest reuse along the loop, and it is suggested to keep it in the innermost position to exploit maximum temporal reuse. 
Similarly, the rest of the loops are ordered based on their partial derivative values.

\subsection{Solving on-chip mapping subspace}
\label{subsec:on-chip-dataflow-search-space}

The on-chip mapping subspace is constructed based on the optimal values of level-3 tile sizes.
Then, our approach explores the constructed subspace to find optimal mappings for each of the three optimal goals, i.e., lower runtime (higher throughput), lower energy consumption, and lower energy-delay product.
For each mapping of the constructed subspace, our approach transforms the mapping into its equivalent MDC notation (described in~\cref{sec:compiler-transformation}).
Then, our approach uses the MDC's cost model~\cite{kwon2018maestro} to estimate various metrics such as latency and energy of each mapping in the on-chip subspace. 
The MDC's cost model precisely computes performance and energy, accounting for under-utilization, edge conditions, and data reuse or movement across time (via L1/L2 buffers~\cite{eyeriss_isca}), space (via broadcast links~\cite{kwon2018maeri}), and space-time (via neighboring links~\cite{jouppi2017datacenter, chen2017eyeriss_issc}) without requiring explicit RTL/cycle-level simulations or access to real hardware.

\begin{algorithm}

    \For{every level-2 inter-tile loop order}{
        \For{every level-2 tile size} {
            Hardware pruning: PE utilization bound \\
            Hardware pruning: No prologues/epilogues \\
            \For{every level-1 tile size} {
            Hardware pruning: Finite L1 size buffer \\
            Hardware pruning: No prologue/epilogue \\
                   // Translate mapping into MDC form \\
                    Invoke the MDC's cost model $\rightarrow$ (runtime, energy, and other metrics)
           }
      }
    }

  \caption{Our approach to explore on-chip mapping subspace, including pruning strategies}
  \label{alg:overview-on-chip-dataflow-search-space}
\end{algorithm}

Algorithm~\ref{alg:overview-on-chip-dataflow-search-space} shows an overview of our approach in exploring the on-chip mapping subspace along with pruning strategies.
We introduce a parameter called ``PE utilization bound ({\tt p})" to prune search space of level-2 tile sizes by bounding the overall PE array utilization to be at-least the parameter {\tt p}. 
The above technique is beneficial in finding optimal on-chip mappings with the optimization goal being throughput, because the highest throughput is typically obtained at higher PE utilization rates~\cite{chen2019eyeriss}.
Our approach also includes a pruning strategy to choose level-1 and level-2 tile sizes such that they don't result in any prologues or epilogues, i.e., the tile sizes are factors of loop bounds.
All of the above-mentioned pruning strategies can be enabled/disabled in \ourtool{}{} by passing them as input parameters.

%% file: tex/04-Evaluation.tex
\section{Evaluation}
\label{sec:evaluation}

In this section, we begin with an overview of the experimental setup used in our evaluation.
Then, we present the evaluation of mappings generated by \ourtool{}{} for a wide variety of DNN operators (CONV2D, GEMM, MLP, and LSTM), and discuss insights from the mappings while comparing them with previous work.

\begin{table}[!ht]
\centering
\scalebox{0.8}{
\begin{tabular}{|c|c|c|}
\hline
\textbf{}                                                               & \textbf{\begin{tabular}[c]{@{}c@{}}Accelerator\\ platform (P1) \\ (Eyeriss-like~\cite{eyeriss_isca})\end{tabular}} & \textbf{\begin{tabular}[c]{@{}c@{}}Accelerator \\ platform (P2) \\ (Edge/IoT-like)~\cite{edge_tpu}\end{tabular}} \\ \hline
\textbf{\#PEs}                                                          & 168                                                                                                               & 1024                                                                                           \\ \hline
\textbf{Clock frequency}                                                & 200 MHz                                                                                                           & 200 MHz                                                                                        \\ \hline
\textbf{GigaOpsPerSec(GOPS)}                                                & 67.2                                                                                                            & 409.6                                                                                        \\ \hline
\textbf{\begin{tabular}[c]{@{}c@{}}NoC bandwidth (GB/s)\end{tabular}} & 2.4                                                                                                               & 25.6                                                                                           \\ \hline
\textbf{L1 buffer size}                                                 & 512B                                                                                                              & 512B                                                                                           \\ \hline
\textbf{L2 buffer size}                                                 & 108KB                                                                                                             & 108KB                                                                                          \\ \hline
\textbf{DRAM block size~\cite{ddr4_spec}}                                                & 64                                                                                                                & 64                                                                                             \\ \hline
\end{tabular}
}
\caption{Accelerator setups in our evaluation.}
\label{tab:hw}
\end{table}

\betterparagraph{Target accelerators} Marvel is applicable to any spatial accelerator since it abstracts accelerator details as \#PEs, L1/L2 buffer sizes, NoC bandwidth, reduction/multicast support, etc, which can be used to model a wide variety of accelerators including Eyeriss~\cite{eyeriss_isca}, NVDLA~\cite{nvdla}, TPU~\cite{edge_tpu}, xDNN. 
Due to space limitations, we present our evaluation for only two accelerator platforms (shown in~\cref{tab:hw}): An accelerator (Eyeriss-like~\cite{eyeriss_isca}) having 168 PEs and 2.4GB/s NoC bandwidth, and another accelerator having 1024 PEs and 25.6GB/s.
We inherit L1, L2 buffer, and clock frequency for both platforms from Eyeriss~\cite{eyeriss_isca}, i.e.,  512B L1 buffer, 108KB L2 buffer, and 200MHz clock frequency.
%
The bidirectional NoC used in our evaluation is a two-level hierarchical bus, which has support for multicasting similar to Eyeriss.
%


\betterparagraph{Experimental variants} We have implemented few of the exploration strategies of recent optimizers such as Interstellar~\cite{interstellar} and dMazeRunner~\cite{DMazeRunner} in our framework.
For, instance, the Interstellar optimizer focuses on parallelizing input and output channels of CONV2D operators, whereas the dMazeRunner optimizer focuses on parallelizing only output channels and a limited set of loop orders.
We compare \ourtool{}{} generated mappings for each workload and accelerator platform with three variants: 1) Marvel implemented Interstellar-like~\cite{interstellar} optimizer generated mappings, 2) Marvel implemented dMazeRunner-like~\cite{DMazeRunner} optimizer generated mappings, and 3) Roof-line peak based on the workload arithmetic intensities and accelerator configurations.

\betterparagraph{Methodology} We have evaluated all the mappings generated by the experimental variants using the MAESTRO cost model~\cite{kwon2018maestro}.
Moreover, the analytical cost model within the MAESTRO framework is validated against the RTL implementations of Eyeriss~\cite{eyeriss_isca} and MAERI~\cite{kwon2018maeri} on VGG16 and AlexNet models.
We passed a pruning option to the \ourtool{}{} to choose tile sizes that divide loop bounds evenly without any remainder, and this has been the consideration in the other approaches~\cite{angshu2019timeloop,interstellar,DMazeRunner,ma2017optimizing,zhang2015optimizing}.
We also set the minimum PE array utilization bound as 0.1, i.e., at-least 10\% of the PE array should be mapped with computation.
We apply 8-bit fixed point precision for all the tensors used in our evaluation.

\subsection{Evaluation on CONV2D}

The CONV2D is a widely used DNN operator in convolution neural networks, and these operators account for more than 90\% of overall computation~\cite{cong2014minimizing, eyeriss_isca}, dominating overall latency, and energy consumption in inferences.
In our evaluation, we considered popular CNN models, such as AlexNet~\cite{Alexnet}, VGG16~\cite{VGGnet}, ResNet50~\cite{Resnet}, and MobileNetV2~\cite{sandler2018mobilenetv2}, with a batch size of one as this captures the low latency requirement use case and also represents a more challenging setup for energy efficiency and throughput~\cite{chen2019eyeriss}.
In addition, these models encompass a broad spectrum of CONV2D operators such as regular, point-wise, depth-wise, strided variants with different filter shapes.

\begin{figure}[!ht]
    \centering
    \includegraphics[width=\linewidth]{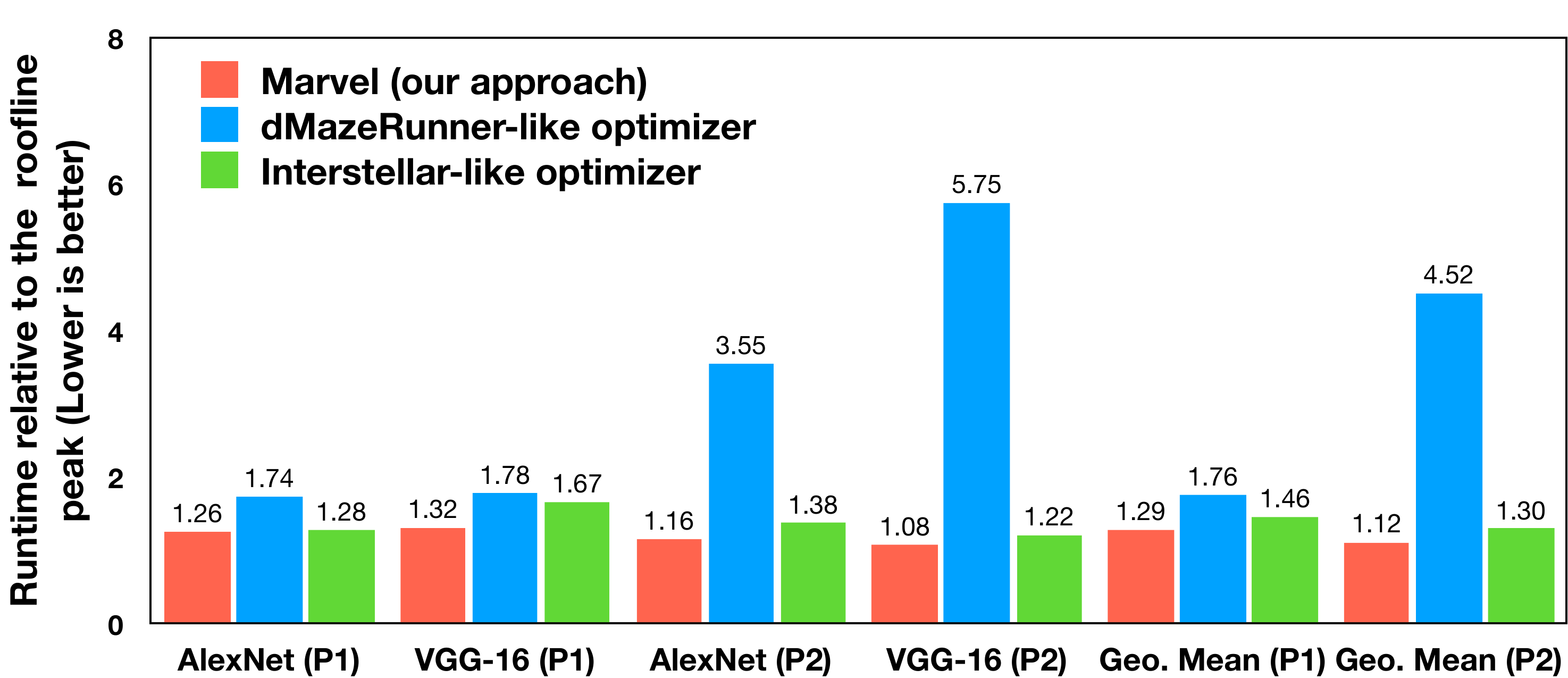}
    \caption{Performance comparison of \ourtool{}{} generated mappings with the mappings of dMazeRunner-like optimizer~\cite{DMazeRunner} and Interstellar-like optimizer~\cite{interstellar} relative to the roof-line peaks of the AlexNet and VGG-16 models on both the platforms (P1 and P2).}
    \label{fig:cnn-results-optimizers}
\end{figure}

\begin{figure*}[!ht]
    \centering
    \includegraphics[width=\linewidth]{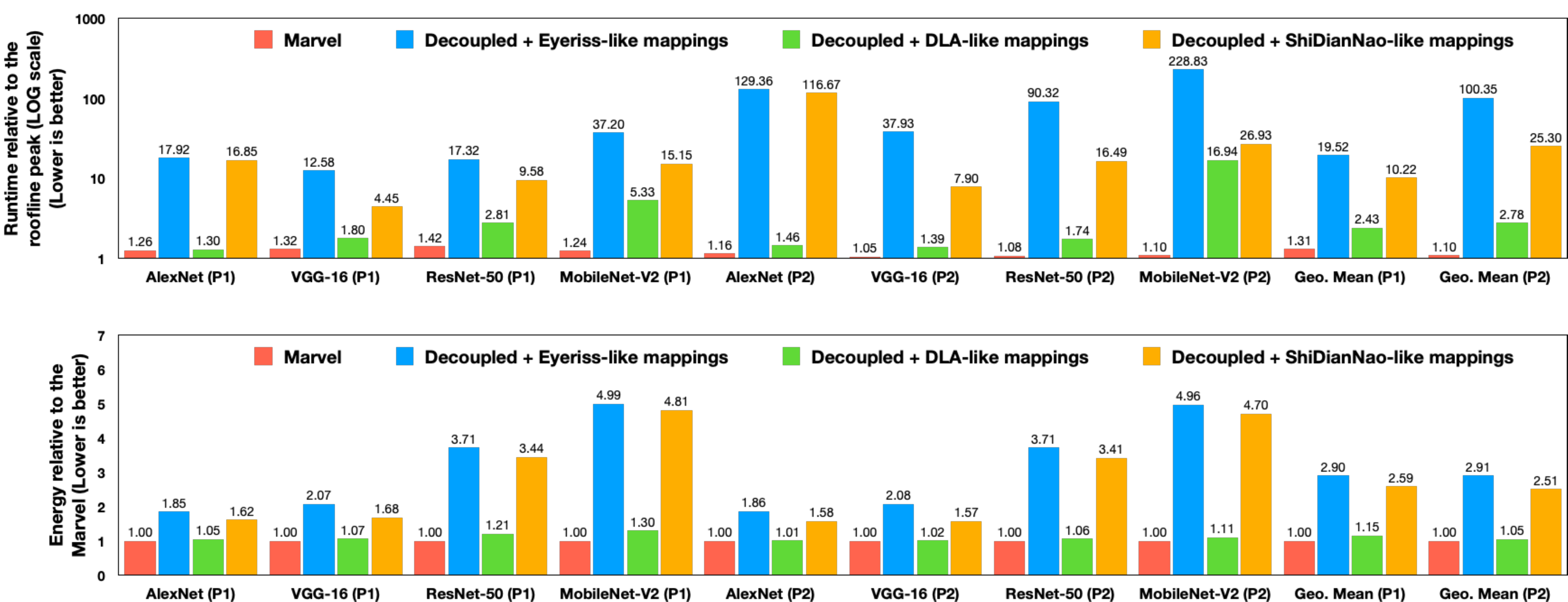}
    \caption{Runtime and energy comparison of \ourtool{}{} generated mappings with the popular mapping styles such as row-stationary (RS) from Eyeriss ~\cite{eyeriss_isca}, weight-stationary from DLA~\cite{nvdla}, output-stationary from ShiDianNao~\cite{du2015shidiannao} for the AlexNet~\cite{Alexnet}, VGG-16~\cite{VGGnet}, ResNet-50~\cite{Resnet}, MobileNet-V2~\cite{sandler2018mobilenetv2} models on both the platforms (P1 and P2).}
    \label{fig:cnn-results}
\end{figure*} 

\betterparagraph{Comparison with the existing optimizers} 
Figure~\ref{fig:cnn-results-optimizers} presents the runtimes of optimized mappings generated by \ourtool{}{}, dMazeRunner-like optimizer~\cite{DMazeRunner}, and Interstellar-like optimizer~\cite{interstellar} relative to the roof-line peaks of the AlexNet and VGG-16 models on both the platforms.
Since each model involves multiple CONV2D operations, we have added the runtimes of the each CONV2D operator to present our evaluation at the level of DNN models.
The Interstellar-like optimizer is almost equivalent to the brute-force exploration except that it restricts exploiting parallelism along only input and output channels. 
As a result, the evaluation using the Interstellar-like optimizer is really time-consuming (multiple days for MobileNetV2 and ResNet50), and hence we restricted the comparison to only AlexNet and VGG16 models.
As can be observed from the~\cref{fig:cnn-results-optimizers},
\ourtool{}{} generated mappings are geometrically 2.35$\times$ and 1.15$\times$ faster compared to the mappings obtained by the dMazeRunner-like optimizer and Interstellar-like optimizer, respectively.
The dMazeRunner-like optimizer focuses on exploiting parallelism along only output channels (in presence of unit batch size) to avoid inter-PE communication, and this results in under-utilization of the PE array for both models.
But, the Interstellar-like optimizer is able to perform close to \ourtool{}{}, because the number of input and output channels in these models are larger (except at the initial layers).
Furthermore, our approach is able to identify mappings in seconds to few minutes for each operator on a local machine, unlike the Interstellar-like optimizer which takes almost 1-5 hours for each operator. 
We don't compare the search time with the dMazeRunner-like optimizer, because we haven't implemented all the heuristic strategies, for, e.g., exploring tiling factors that highly utilize (at-least 60 \%) the scratchpad buffers.
Table~\ref{tab:search_space} shows the impact of our decoupling and pruning strategies on the original search space of mappings of the four models with an average reduction of $O(10^{10})$ in the mapping space.
\input{tables/search_space_table.tex}

\begin{figure*}[!ht]
    \centering
    \includegraphics[width=\linewidth]{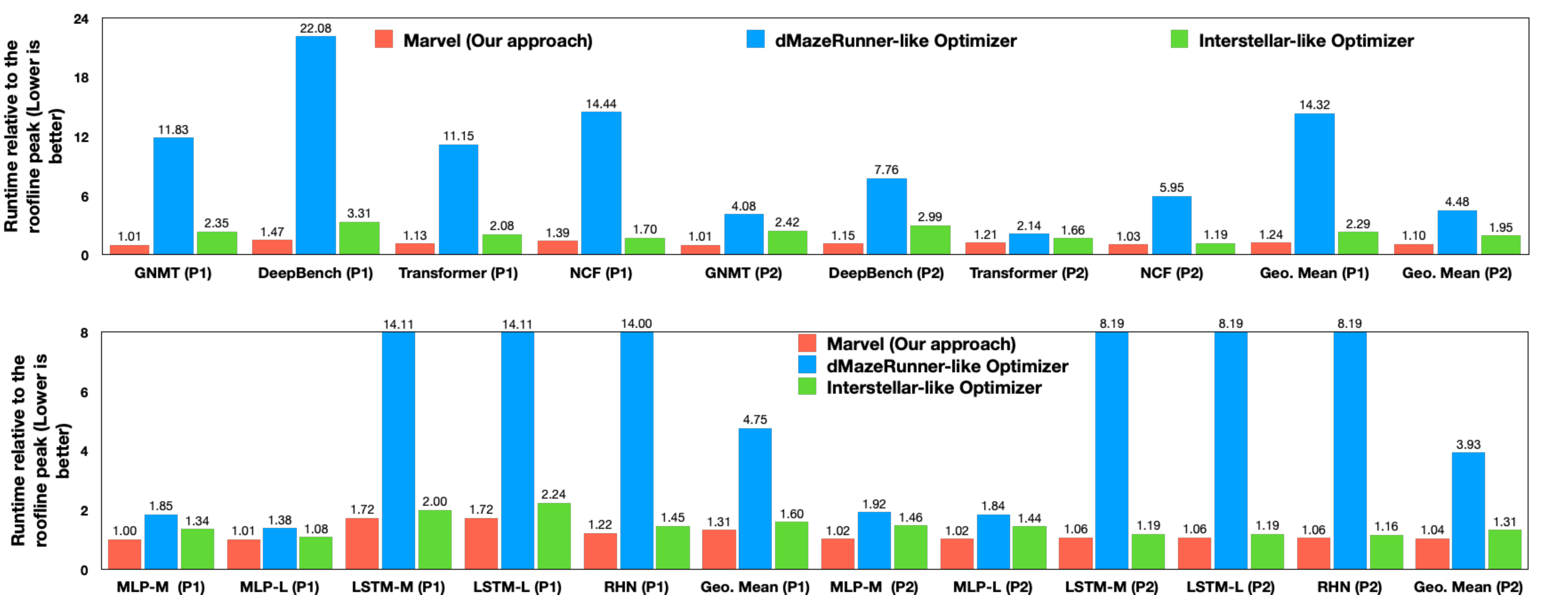}
    \caption{Performance comparison of \ourtool{}{} generated mappings with the mappings of dMazeRunner-like optimizer~\cite{DMazeRunner}, and Interstellar-like optimizer~\cite{interstellar} relative to the roof-line peaks of the GEMM workloads in~\cref{tab:gemm-workloads} and LSTM, MLP in~\cref{tab:mlp-size} on both the platforms (P1 and P2).}
    \label{fig:gemm-lstm-results}
\end{figure*}



\betterparagraph{Comparison with the popular mapping styles} Some of the state-of-the-art mapping styles are row-stationary (RS) from Eyeriss ~\cite{eyeriss_isca}, weight-stationary from DLA~\cite{nvdla}, and output-stationary from ShiDianNao~\cite{du2015shidiannao}.
In our evaluation, we encoded the above mapping styles in the form of parallelization and loop order constraints on the on-chip mapping space of our decoupled approach. 
For instance, weight-stationary (DLA) mapping style includes parallelization across input and output channels with the loop iterators corresponding to the weight tensor in the innermost positions of the loop orders.
As can be observed from~\cref{fig:cnn-results}, the runtimes of \ourtool{}{} generated mappings for all the models are only 1.31$\times$ and 1.10$\times$  higher relative to the roof-line peaks of all the models on both accelerator platforms P1 and P2, respectively.

The Eyeriss-like mappings~\cite{eyeriss_isca} exploit parallelism along output width and filter width dimensions, where as the ShiDianNao-like mappings~\cite{du2015shidiannao} exploit along output width and height.
But, the extents of these dimensions are relatively small especially in modern DNN models such as ResNet50 and MobileNetV2.
Hence, these mappings are often resulted in under-utilization of the PE array leading to higher runtimes compared to the roof-line peak (e.g., 100.36$\times$ for Eyeriss-like mappings on platform P2).
But, these mappings exploit popular row-stationary and output-stationary behavior leading to lower energy consumption (e.g., 2.91$\times$ for Eyeriss-like mappings on platform P2) relative to the \ourtool{}{} reported energy-efficient mappings.

The DLA-like mappings exploit parallelism along input and output channels, and the extent of these dimensions are sufficient enough to keep the PE array busy for most of the layers of AlexNet, VGG16, and ResNet50 models. 
However, the MobileNetV2 model has introduced depth-wise operators which lacks parallelism in the input channels.
This resulted in less performance of the DLA-like mapping compared to the roof-line peak, and our approach exploited alternate dimensions (more than one) for the parallelism.
However, the DLA-like mappings exploit weight-stationary reuse behavior, and these DNN models have large number of weight parameters compared to other tensors. This resulted in only 1.10$\times$ higher energy consumption relative to the \ourtool{}{} reported energy-efficient mappings.

\subsection{Evaluation on GEMM}

In this evaluation, we have considered GEMM workloads from the recent work in~\cite{qin2020sigma}.
An interesting aspect of these workloads is that they are irregular in their shapes making the rigid accelerators (e.g., TPUs) hard to reach their peak utilization~\cite{qin2020sigma}.
A summary of these workloads are shown in~\cref{tab:gemm-workloads}, where M, N, K refers to number of rows, columns of first matrix followed by the columns of second matrix.

\begin{table}[!ht]
\scalebox{0.9}{
\begin{tabular}{|c|c|c|c|c|}
\hline
\multirow{2}{*}{\textbf{Workload}}    & \multirow{2}{*}{\textbf{Application}}                                                       & \multicolumn{3}{c|}{\textbf{Dimensions}} \\ \cline{3-5} 
                                      &                                                                                             & \textbf{M}   & \textbf{N}  & \textbf{K}  \\ \hline
\multirow{4}{*}{\textbf{GNMT}}        & \multirow{4}{*}{\textbf{\begin{tabular}[c]{@{}c@{}}Machine \\ Translation\end{tabular}}}    & 128          & 2048        & 4096        \\ \cline{3-5} 
                                      &                                                                                             & 320          & 3072        & 4096        \\ \cline{3-5} 
                                      &                                                                                             & 1632         & 36548       & 1024        \\ \cline{3-5} 
                                      &                                                                                             & 2048         & 4096        & 32          \\ \hline
\multirow{2}{*}{\textbf{DeepBench}}   & \multirow{2}{*}{\textbf{\begin{tabular}[c]{@{}c@{}}General\\ Workload\end{tabular}}}        & 1024         & 16          & 500000      \\ \cline{3-5} 
                                      &                                                                                             & 35           & 8457        & 2560        \\ \hline
\multirow{2}{*}{\textbf{Transformer}} & \multirow{2}{*}{\textbf{\begin{tabular}[c]{@{}c@{}}Language\\ Understanding\end{tabular}}}  & 31999        & 1024        & 84          \\ \cline{3-5} 
                                      &                                                                                             & 84           & 1024        & 84          \\ \hline
\multirow{2}{*}{\textbf{NCF}}         & \multirow{2}{*}{\textbf{\begin{tabular}[c]{@{}c@{}}Collaborative\\ Filtering\end{tabular}}} & 2048         & 1           & 128         \\ \cline{3-5} 
                                      &                                                                                             & 256          & 256         & 2048        \\ \hline
\end{tabular}
}
\caption{Description of the GEMM workloads which are taken from the recent work in~\cite{qin2020sigma}.}
\label{tab:gemm-workloads}
\end{table}

We translated the GEMM workloads into their equivalent CONV2D workloads for the Interstellar-like and dMazeRunner-like optimizers, because their exploration strategies are specific to the CONV2D workloads (e.g., parallelization strategies).
Figure~\ref{fig:gemm-lstm-results} presents the runtime of optimized mappings generated by \ourtool{}{}, dMazeRunner-like optimizer~\cite{DMazeRunner}, and Interstellar-like optimizer~\cite{interstellar} relative to the roof-line peak of each GEMM workload.
The runtimes of \ourtool{}{} generated mappings are only 1.24$\times$ and 1.10$\times$ higher relative to the roof-line peaks of accelerator platforms P1 and P2 respectively, thereby demonstrating the closeness of mappings obtained using our approach to the peak.
Furthermore, we observed that maximum reuse (spatial, temporal, spatio-temporal) is exploited only when all the dimensions of the GEMM operator are parallelized. 
Hence, \ourtool{}{} generated mappings included parallelization of the three dimensions to make the PE array occupied along with exploiting maximum reuse.
This is in contrast to other approaches, i.e., Interstellar-like optimizer focusing on parallelizing only (N,K) dimensions and dMazeRunner-like optimizer focusing on parallelizing only (K) dimension.
 As a result, \ourtool{}{} generated mappings are 6.87$\times$ and 1.81$\times$ faster compared to the mappings obtained by the dMazeRunner-like optimizer and Interstellar-like optimizer for all the GEMM workloads on the both accelerator platforms.

\begin{figure*}[!ht]
    \centering
    \includegraphics[width=\linewidth]{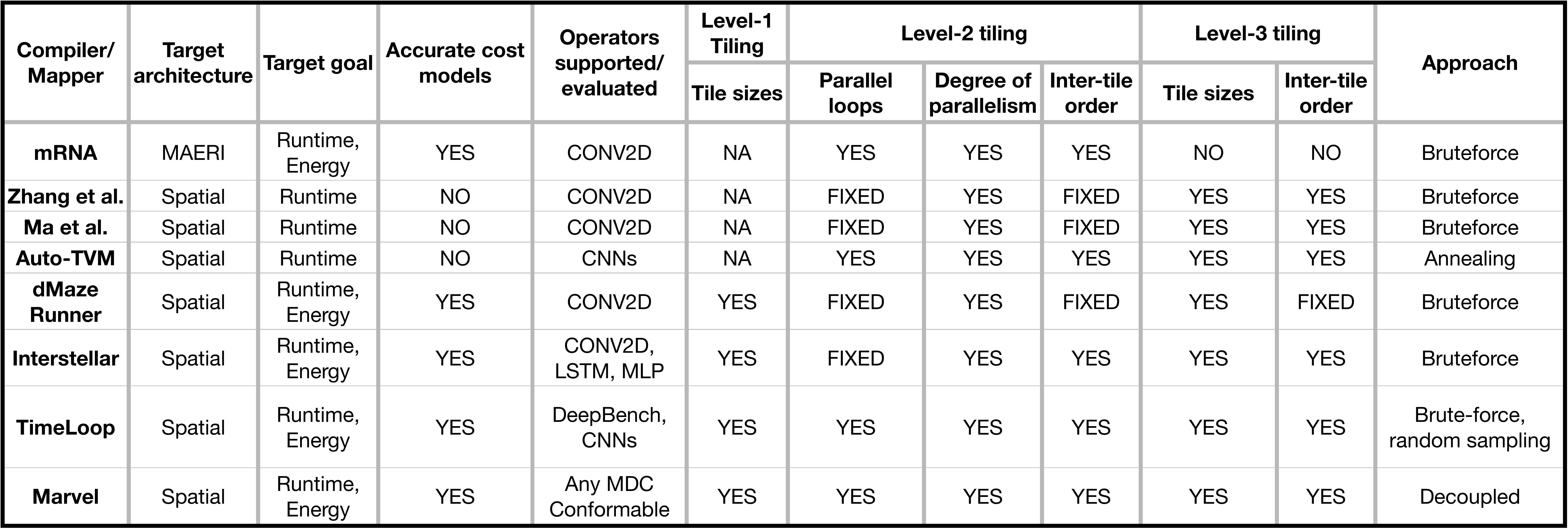}
    \caption{Comparison of \ourtool{}{} with prior approaches (mRNA~\cite{zhao2019mRNA}, Zhang et al.~\cite{zhang2015optimizing}, Ma et al.~\cite{ma2017optimizing}, Auto-TVM~\cite{Chen:2018:TAE:3291168.3291211}, dMazeRunner~\cite{DMazeRunner}, Interstellar~\cite{interstellar}, TimeLoop~\cite{angshu2019timeloop}) for the mapping space exploration of DNN operators. Our approach (\ourtool{}{}) supports any operator conformable with the MDC notation.}
    \label{fig:priorwork}
\end{figure*}

\subsection{Evaluation on MLP and LSTM}

In this evaluation, we have considered the MLP and LSTM workloads from the Interstellar work in~\cite{interstellar}, and a summary of these workloads are shown in~\cref{tab:mlp-size}.

\begin{table}[!ht]
\begin{tabular}{|c|c|c|c|}
\hline
\textbf{Network}                & \textbf{Layer}   & \textbf{Input channels}   & \textbf{Output channels} \\ \hline
\multirow{3}{*}{\textbf{MLP-M}} & \textbf{FC1}     & 784                       & 1000                     \\ \cline{2-4} 
                                & \textbf{FC2}     & 1000                      & 500                      \\ \cline{2-4} 
                                & \textbf{FC3}     & 500                       & 250                      \\ \hline
\multirow{3}{*}{\textbf{MLP-L}} & \textbf{FC1}     & 784                       & 1500                     \\ \cline{2-4} 
                                & \textbf{FC2}     & 1500                      & 1000                     \\ \cline{2-4} 
                                & \textbf{FC3}     & 1000                      & 500                      \\ \hline
                                \hline
\textbf{Network}                & \multicolumn{2}{c|}{\textbf{Embedding size}} & \textbf{Batch size}      \\ \hline
\textbf{LSTM-M}                 & \multicolumn{2}{c|}{500}                     & 128                      \\ \hline
\textbf{LSTM-L}                 & \multicolumn{2}{c|}{1000}                    & 128                      \\ \hline
\textbf{RHN}                    & \multicolumn{2}{c|}{1500}                    & 128                      \\ \hline
\end{tabular}
\caption{Description of the MLP and LSTM workloads which are  taken from the Interstellar work in~\cite{interstellar}.}
\label{tab:mlp-size}
\end{table}

We translated the MLP workloads into CONV2D workloads for the Interstellar-like and dMazeRunner-like optimizers.
We also translated the LSTMs workloads into their equivalent CONV2D workloads via first converting into GEMM workloads.
For instance, a LSTM workload with batch size as {\tt B} and embedding size\footnote{Embedding size is the size of input and hidden vectors.} as {\tt E} can be translated into a GEMM workload with M being the batch size (B), N being the embedding size (E), and K being the 2$\times$E.
Figure~\ref{fig:gemm-lstm-results} presents the runtime of optimized mappings generated by \ourtool{}{}, dMazeRunner-like optimizer~\cite{DMazeRunner}, and Interstellar-like optimizer~\cite{interstellar} relative to the roof-line peak of each workload in~\cref{tab:mlp-size}.
\ourtool{}{} generated mappings are 4.46$\times$ and 1.22$\times$ faster compared to the mappings obtained by the dMazeRunner-like optimizer and Interstellar-like optimizer for all the workloads on the both accelerator platforms.
The benefits compared to dMazeRunner-like optimizer is higher because of its parallelization across only a single dimension (Embedding size in case of LSTM and Output channels in case of MLP) and also exploring only limited loop orders for reuse. 
In addition, \ourtool{}{} is able to do better compared to Interstellar-like optimizer by exploring more levels of parallelism to make the PE array occupied (e.g., only 1.04$\times$ higher relative to roof-line peak on platform P2).

%% file: tables/search_space_table.tex
\begin{table}[!ht]
 \centering
\scalebox{0.8}{
\begin{tabular}{|l|l|l|l|}
\hline
\multicolumn{1}{|c|}{\multirow{2}{*}{Variants}} & \multicolumn{3}{c|}{\textbf{Search space size}} \\ \cline{2-4} 
\multicolumn{1}{|c|}{} & \multicolumn{1}{c|}{\textbf{Min}} & \multicolumn{1}{c|}{\textbf{Avg}} & \multicolumn{1}{c|}{\textbf{Max}} \\ \hline
Original search space & \multirow{2}{*}{2.7$\times 10^{17}$} & \multirow{2}{*}{9.4$\times 10^{18}$} & \multirow{2}{*}{1.8$\times 10^{19}$} \\ 
 &  &  &  \\ \hline
Off-chip schedules search & \multirow{2}{*}{7.3$\times 10^{8}$} & \multirow{2}{*}{3.6$\times 10^{11}$} & \multirow{2}{*}{1.3$\times 10^{12}$} \\ 
space after decoupling &  &  &  \\ \hline
On-chip schedules search & \multirow{2}{*}{2.9$\times 10^{7}$} & \multirow{2}{*}{2.4$\times 10^{10}$} & \multirow{2}{*}{1.4$\times 10^{11}$} \\ 
space after decoupling &  &  &  \\ \hline
Off-chip schedules search & \multirow{2}{*}{9.9$\times 10^{5}$} & \multirow{2}{*}{1.5$\times 10^{8}$} & \multirow{2}{*}{6.3$\times 10^{8}$} \\ 
space after decoupling + pruning &  &  &  \\ \hline
On-chip schedules search & \multirow{2}{*}{3.8$\times 10^{5}$} & \multirow{2}{*}{5.9$\times 10^{7}$} & \multirow{2}{*}{2.4$\times 10^{8}$} \\ 
space after decoupling + pruning &  &  &  \\ \hline
\end{tabular}
}
\caption{The statistics (min/avg/max) of the CONV2D mapping space in our evaluation and the resultant mapping subspaces after decoupling and pruning strategies.}
\label{tab:search_space}
\end{table}

%% file: tex/05-Relatedwork.tex
\vspace{-2mm}
\section{Related Work}
\label{sec:related}

In this section, we discuss prior work only on compilers/mappers (shown in Figure~\ref{fig:priorwork}) for finding efficient mappings of DNN operators on to the spatial accelerators.
Prior work~\cite{zhao2019mRNA,eyeriss_isca} focused on developing mappers specific to their architectures, for, e.g., mRNA mapper~\cite{zhao2019mRNA} for the MAERI accelerator~\cite{kwon2018maeri}, limiting their applicability to generic spatial accelerators.
Prior work such as Auto-TVM~\cite{Chen:2018:TAE:3291168.3291211}, Zhang et al.~\cite{zhang2015optimizing}, Ma et al.~\cite{ma2017optimizing} focused on spatial accelerators without L1 buffers inside a PE, again limiting their mapping space formulation.
Furthermore, they don't employ accurate cost models and focus only on optimizing for runtime. 

In addition, other prior works such as Interstellar~\cite{interstellar}, dMazeRunner~\cite{DMazeRunner} fixed certain aspects of mapping space such as choice of parallel loops, loop orders, and these choices may not reflect the efficient mappings for a wide variety of DNN operators.
To the best of our knowledge, TimeLoop~\cite{angshu2019timeloop} is the only framework that considers all aspects of a mapping for a fully flexible spatial accelerator. 
However, it employs either an exhaustive linear search or a random sampling-based heuristic to explore the search space.
In contrast to all of the above works, our approach considers all the aspects of mapping space and uses the decoupled strategy to efficiency navigate the mapping space.

Most of the prior work focus on optimizing convolution operators, and its not clearer if their approaches are applicable to any DNN operator expressed in the loop-nest form. But, our approach is guaranteed to perform on any operator conformable to the MDC notation.

%% file: tex/06-Conclusion.tex
\section{Conclusion \& Future work}
\label{sec:conclusion}

In this paper, we provide a formal understanding of DNN operators whose mappings can be described in the MDC notation by introducing a set of rules over the abstract loop nest form of the operators.
Furthermore, we introduce a transformation for translating mappings into the MDC notation for exploring the mapping space.
Then, we also proposed a decoupled off-chip/on-chip approach that decomposes the mapping space into off-chip and on-chip subspaces, and first optimizes the off-chip subspace followed by the on-chip subspace. 
We implemented our decoupled approach in a tool called {\em Marvel}, and a major benefit of our approach is that it is applicable to any DNN operator conformable with the MDC notation.
Our approach reduced the search space of CONV2D operators from four major DNN models from $9.4 \times 10^{18}$ to $1.5 \times 10^{8} + 5.9 \times 10^{8} \approxeq 2.1 \times 10^8$, which is a reduction factor of ten billion (Table~\ref{tab:search_space}), while generating mappings that demonstrate a geometric mean performance improvement of 10.25$\times$ higher throughput and 2.01$\times$ lower energy consumption compared with three state-of-the-art mapping styles from past work.
In the future, we envision 1) \ourtool{} integration with the MLIR compiler infrastructure for wide usability, 2) extending the MDC notation and its cost model to support non-conformable operators, and also 3) using for a wide range of applications, including the neuro-architecture search.


